# Edwin P. Hubble, astropolítico

## Martín López Corredoira

*1. Vida*

Pocas cosas se pueden decir de la biografía de un científico ampliamente reconocido que no se puedan encontrar hoy en día en una búsqueda rápida por Internet, y me temo que no será ésta una excepción. No obstante, más que ofrecer un conocimiento enciclopédico cronológicamente ordenado de toda una vida, intentaré señalar solamente algunos de los eventos más llamativos de la existencia del personaje, seleccionados de algunas notas biográficas.[1]

Edwin Powell Hubble nació en Marshfield (Misuri, Estados Unidos) el 20 de noviembre de 1889. Su padre trabajaba en una compañía de seguros y quizá por su cercanía a cuestiones legales tuvo la idea de empujar a Edwin a estudiar derecho, sin mucho interés por parte del hijo. Sin embargo, un año después de fallecer su progenitor (en 1913), nuestro protagonista da un giro profesional en su carrera hacia la astronomía, afición que ya le venía de joven. En cierta ocasión, en la correspondencia que mantenía con uno de sus abuelos, el joven Edwin de 12 años envió una respuesta acerca de cuestiones sobre el planeta Marte que sorprendió al abuelo, el cual decidió publicarla en un periódico. Se convertiría luego en la figura más notable de la astronomía norteamericana, y trabajaría en los más importantes observatorios astronómicos de Estados Unidos, Monte Wilson y Monte Palomar.

Estudió también español y llegó a ser profesor de lengua española, junto con otras asignaturas como Física y Matemáticas, en la New Albany High School (New Albany, Indiana, Estados Unidos) en 1913-1914. De Hubble se sabe también de su destreza e interés como deportista, como atleta o boxeador o jugador de diversos deportes de pelota. Participó en ambas Guerras Mundiales, llegando a ser comandante (*major*) durante la primera contienda.

Se casó en California en 1924 con Grace Burke, con quien permaneció unido hasta el final de su vida, el 28 de septiembre de 1953, en San Marino (California). No se celebró ninguna ceremonia de funeral, y su esposa nunca revelaría el lugar donde fue enterrado.

*2. Hombre de leyes*

Aunque llegaría a ser conocido nuestro protagonista por descubrir leyes del Universo, no le era ajeno el mundo de las leyes de nuestra sociedad civil, el mundo del derecho. Conocía su lenguaje por su experiencia como becario entre 1910 y 1913 estudiando derechos romano e inglés en el Queen College de Oxford (Reino Unido), donde fue hecho miembro honorario. A su retorno a Estados Unidos en 1913, pasó los exámenes que le permitían ejercer en una jurisdicción (*bar examination*) y tuvo un año de prácticas como abogado en Louisville (Kentucky, Estados Unidos).[2]

---

[1] Mayall, 1970; https://en.wikipedia.org/wiki/Edwin_Hubble ;
https://www.spacetelescope.org/about/history/the_man_behind_the_name/

*3. Interés por la filosofía*

En el mundo anglosajón, se considera a Hubble un hombre culto en aspectos humanistas, principalmente por su interés por la historia y la filosofía de la ciencia. Para un erudito intelectual europeo de su época, quizá no fueran muy profundos sus intereses filosóficos, sin sacar los pies del tiesto de la ciencia, pero para la tradición analítica y especializada de su país, esto era bastante. Se conoce por alguno de sus manuscritos que dedicó algún tiempo a estudiar la ciencia inglesa en el renacimiento y a Francis Bacon.[3] Escribió además diversos ensayos filosóficos sobre la naturaleza de la ciencia,[4] mostrándose riguroso metodológicamente y a favor de una interpretación de experimentos y observaciones sin prejuicios basados en principios a priori o en valores estéticos o juicios personales. Abogaba por una ciencia moderna experimental que superara la dialéctica medieval.

Sin embargo, sus principios sobre metodología de las ciencias no fueron precisamente los que aplicó a su trabajo como científico:[5] a la hora de confrontar los principios generales con los datos observacionales, prefirió dar más peso a los primeros. Los principios de uniformidad de la naturaleza le llevaron a descartar modelos cosmológicos como el de Milne, que explicaban perfectamente los datos extragalácticos. Hubble prefirió la teoría establecida a priori, antes que derivar inductivamente cómo debe ser el modelo teórico. Hubble sería no obstante también reacio a aceptar el consenso del modelo relativista en un Universo en expansión hasta por lo menos el año 1937.[6] En su trabajo científico, afirmaba que la teoría comienza con ciertos principios fundamentales que ayudarían a seleccionar el rango de los mundos lógicamente posibles a comparar con las observaciones,[7] y esto choca con el Hubble filósofo empirista.

*4. Astrónomo, astrofísico, cosmólogo*

El brillante Hubble tendría un gran futuro como astrónomo o astrofísico observacional, en particular dentro del área de "galaxias y cosmología". Es considerado uno de los padres de la astronomía extragaláctica, es decir de los fenómenos fuera de nuestra Galaxia la Vía Láctea, aunque, por su enemistad con el también astrónomo norteamericano Harlow Shapley, nunca empleó la palabra "galaxia" propuesta por aquel, sino el término "nebulosa extragaláctica" (*extragalactic nebula*).[8] Además, sus trabajos sobre otras galaxias darían soporte a la actual teoría estándar de cosmología.

---

[2] Mayall, 1970.
[3] Hetherington, 1982.
[4] E.P. Hubble, *The nature of science and other lectures*, San Marino (California, EE.UU.), Huntington library, 1954.
[5] Hetherington, 1982.
[6] Gale, 1993.
[7] Hetherington, 1982, secc. 3.

No obstante, no es oro todo lo que reluce y el brillo deslumbrante de la figura de Hubble tiene su leyenda negra. Todos o casi todos sus grandes descubrimientos fueron repeticiones de hallazgos de otros investigadores que quedaron a la sombra de la superestrella norteamericana en la injusta Historia, y a día de hoy se está revisando el valor innovador de su ciencia. El esquema de clasificación de galaxias, el llamado diapasón (*tuning fork*) de Hubble, los perfiles de luminosidad de galaxias, etc. fueron ideas desarrolladas por otros astrofísicos con anterioridad.[9] Las distancias a galaxias habían sido también medidas por otros astrónomos. Por ejemplo, la distancia de la galaxia de Andrómeda había sido medida por el astrónomo estonio Ernst Julius Öpik o por el astrónomo sueco Knut Lundmark antes que Hubble en 1925, aunque sólo después de las investigaciones de Hubble sería aceptada la medida por la comunidad a pesar de usar un método más impreciso utilizando estrellas cefeidas como calibradores.[10] Tampoco, como veremos en la próxima sección, son originalmente suyos los hallazgos del desplazamiento al rojo de las galaxias, su correlación con la distancia en la conocida como ley de Hubble o su interpretación en términos de la expansión del Universo.

## 5. El descubrimiento de la expansión del Universo

Hay antecedentes previos a la proclamación del descubrimiento de la ley de Hubble en 1929,[11] la cual relaciona linealmente la recesión de galaxias con su distancia. El astrónomo alemán Carl Wirtz ya se dio cuenta en 1924 de la anticorrelación entre el brillo de una galaxia (y cuento más lejos está, con menos brillo es observada) y su desplazamiento espectral al rojo.[12] Tampoco los datos son totalmente propios: Hubble utiliza en 1929 los desplazamientos al rojo que habían sido medidos por el astrónomo estadounidense Vesto Melvin Slipher, publicados por el astrofísico británico Arthur Stanley Eddington en 1923 y añadidos a los estudios de Milton Humason, colaborador de Hubble, en 1929 antes del anuncio de su descubrimiento.[13] Si bien, las medidas de las distancias de las galaxias son mayormente de su equipo.

---

[8] Madrid Casado, 2018, cap. 19.
[9] Block, 2011; Way, 2013.
[10] Einasto, 2001.
[11] Hubble, 1929.
[12] En física y astronomía, el desplazamiento hacia el rojo (en inglés: *redshift*) ocurre cuando la radiación electromagnética que se emite desde un objeto es desplazada hacia menores frecuencias (hacia el rojo en luz visible). Un desplazamiento hacia el rojo puede ocurrir cuando una fuente de luz se aleja de un observador, correspondiéndose a un desplazamiento Doppler que cambia la frecuencia percibida de las ondas. La espectroscopia astronómica utiliza los desplazamientos al rojo Doppler para determinar el movimiento de objetos astronómicos distantes. Otro mecanismo de corrimiento hacia el rojo es la expansión métrica del espacio, que explica la observación de los desplazamientos al rojo espectrales de galaxias distantes se incrementan proporcionalmente con su distancia al observador. Este mecanismo es una característica clave del modelo del Big Bang de la cosmología física (https://es.wikipedia.org/wiki/Corrimiento_al_rojo).
[13] Eddington, 1923; Humason, 1929.

La ley de Hubble ha pasado a llamarse desde 2018 por recomendación de la IAU (Unión Astronómica Internacional) ley de Hubble-Lemaître, con el fin de hacer justicia a su descubridor original. El sacerdote católico, matemático, astrónomo y físico belga Georges Lemaître había publicado en 1927 un trabajo en francés en una desconocida revista belga, describiendo tanto su modelo teórico de universo como un análisis de datos de galaxias que muestran tal ley. Algunos estudiosos de la historia del descubrimiento sugieren que Hubble encontró justo lo que andaba buscando en el trabajo de Lemaître. Su colaborador Humason relató en una entrevista concedida en 1965 que Hubble supo de la relación velocidad-distancia por la comunicación presentada por Lemaître en una reunión de la IAU en Holanda en 1928,[14] aunque es probable que no leyera su artículo de 1927 por estar escrito en francés. Como quiera que sea, en 1929, Hubble y Humason publican su descubrimiento de la ley, sin citar el trabajo de Lemaître y se queda Hubble todo el mérito del descubrimiento. En 1931, Lemaître tendría la oportunidad de traducir su artículo de 1927 al inglés y publicarlo en una revista británica de alto prestigio y difusión, *Monthly Notices of the Royal Astronomical Society*, pero cometió la torpeza de eliminar párrafos y fórmulas referentes al análisis de datos que mostraban que él había descubierto la recesión de galaxias y su relación lineal con la distancia antes que Hubble,[15] así que la contribución de Lemaître permanecería desconocida por varias décadas más ante el Statu Quo de la cosmología angloparlante.

Hoy además sabemos que las 24 galaxias que utilizó Hubble para proclamar su famosa ley relacionada con la expansión del Universo estaban demasiado cerca y dominadas por movimientos propios. Fue una coincidencia que se pudiera ver una relación lineal en estos datos. Los datos en 1929 no eran suficientemente buenos para poder concluir la ley de Hubble-Lemaître, lo que muestra cómo los prejuicios teóricos guiaron la investigación.[16]

### *6. Duda razonable sobre la expansión*

El abogado del diablo y el filósofo de dudas metódicas que habían en Hubble contribuyeron probablemente a alimentar el escepticismo sobre sus propios logros, algo muy saludable para una ciencia abierta y no-dogmática, al contrario de lo que sucedería entre los cosmólogos posteriores a la muerte de Hubble, a quienes describió bastante certeramente el Premio Nobel de Física Lev Landáu cuando exclamó: "Los cosmólogos cometen errores a menudo, pero nunca dudan".

Hubble nunca llegó a estar del todo satisfecho con la idea del Universo en expansión.[17] En el propio artículo de 1929 sobre el descubrimiento de su ley afirma: "Los nuevos datos que se esperan en un futuro pueden modificar el nivel al que es significativo el resultado de la presente investigación, o dando mayor peso a la solución si se confirma. Por esta razón, se cree que es prematuro discutir en detalle las consecuencias obvias de los presentes resultados".[18] Era consciente de que sus datos no eran suficientemente conclusivos, como he señalado anteriormente. Pero incluso muchos años después tampoco llegó a convencerse del todo de la explicación de la expansión, a pesar de que la calidad de los datos había mejorado. Tenía en mente que podría haber otra explicación alternativa a la expansión para explicar el desplazamiento al rojo de los fotones, la llamada hipótesis del "fotón cansado", aunque no le diese credibilidad alguna con los datos conseguidos hasta aquel entonces: "los fotones emitidos por una nébula pierden energía en su viaje al observador por algún efecto desconocido, el cual es lineal con la distancia y conlleva una disminución en frecuencia sin una desviación transversal en su trayectoria apreciable y, en particular, sin disminución alguna de la proporción de fotones llegada al observador (…) parece posible que los desplazamientos al rojo no sean debidos a la expansión del Universo, y puede requerirse una reexaminación de mucha de la especulación sobre la estructura del Universo".[19]

---

[14] Llallena Rojo, 2017, p. 90.
[15] Livio, 2011.
[16] Bonometto, 2001.
[17] Eichler, 2015; Sandage, 1989, p. 357.

Hubble[20] se percata también de que, en su modelo de expansión del Universo con la constante $H_0$ que había medido (se supo posteriormente que tal medida fue incorrecta, por un problema de calibración de distancias), el Universo debería tener una edad de menos de dos mil millones de años, lo que resulta ser desconcertantemente joven, menor que la edad de la Tierra, además de resultar un universo demasiado denso. Cabe la posibilidad, según Hubble, de que la interpretación de los desplazamientos de las líneas espectrales al rojo no tenga que ver con la recesión de las galaxias, independientemente de que el Universo se expanda o contraiga a algún ritmo diferente del medido si es que lo hace. Esto le lleva a optar "como sucediera una vez en los tiempos de Copérnico, entre un universo finito, y un universo sensiblemente infinito más un nuevo principio de la naturaleza. Y, como antaño, la elección puede venir determinada por el atributo de simplicidad".[21]

## 7. *El astropolítico*

Hubble fue ante todo un gran astropolítico. uno de los maestros de la tradición de astropolítica actual. ¿Qué es un astropolítico? Es un líder de un grupo de investigación en astronomía o astrofísica que destaca como gestor o administrador de proyectos científicos, y que en vez de centrarse en su labor científica se dedica más bien a hacer vida social, buscar financiación y estar al tanto de las ideas de los demás para luego escoger las que le interesan y poner a sus súbditos a trabajar en el tema, haciendo suyo el descubrimiento una vez se pone en lugar visible destacado, tras obtener su grupo resultados y promoverlos con la campaña de

---

[18] "New data to be expected in the future may modify the significance of the present investigation or, if confirmatory, will lead to a solution having many times the weight. For this reason, it is thought premature to discuss in detail the obvious consequences of the present results." (Hubble, 1929).

[19] "(…) the photons emitted by a nebula lose energy on their journey to the observer by some unknown effect, which is linear with distance and which leads to a decrease in frequency without appreciable transverse deflection and, in particular, without any decrease in rate of arrival at the observer (…) it seems likely that red-shifts may not be due to an expanding Universe, and much of the speculation on the structure of the universe may require re-examination" (Hubble, 1947).

[20] Hubble, 1942.

[21] "(…) as once before in the days of Copernicus, between a small, finite universe, and a sensibly infinite universe plus a new principle of nature. And, as before, the choice may be determined by the attribute of simplicity" (Hubble, 1942, p. 215).

*marketing* en medios de comunicación y a través de la influencia en organización de congresos.[22] Esta figura surge en una época en que la ciencia precisa de grandes inversiones económicas para poder avanzar, en particular en telescopios gigantes en astronomía, y la figura del hábil gestor que convence a políticos y donantes privados se hace fundamental y hasta más importante que la del investigador enclaustrado en su telescopio afanado por poner su inteligencia y su tesón al servicio del conocimiento del espacio. El astropolítico es el triunfador de nuestro tiempo en astronomía y astrofísica, el que se lleva los laureles dentro de la maquinaria del sistema científico;[23] por supuesto, lo mismo es extrapolable a otras ciencias en la actualidad. Hubble, como pionero de la astropolítica, no tuvo el grado de desarrollo en tal competencia que tienen los profesionales hoy en día, a cuyo lado pasaría como un simple aficionado, pues todavía era de los que iba a observar a los telescopios y hacía parte del trabajo manual. No obstante, puso las bases sobre cómo medrar en una carrera científica hasta lo más alto sin destacar por aportar descubrimientos importantes y *originales*, lo que sería un modelo a imitar y mejorar por nuestros administradores de la ciencia actuales ávidos de poder e influencia.

La faceta ambiciosa se vio ya en Hubble ante el descubrimiento de la expansión del Universo y la ley de Hubble-Lemaître. Seguramente Hubble sabía o sospechaba que no había sido el primero en postular tal, pero necesitaba un gran éxito en su carrera para él y su equipo a fin de justificar el enorme gasto por aquella época del telescopio Hooker de 2,5 metros en el Observatorio del Monte Wilson. En una carta dirigida al cosmólogo holandés Willem de Sitter que escribiría el 21 de agosto de 1930 afirmaría: "Considero la relación velocidad-distancia, su formulación, comprobación y confirmación, como una contribución de Monte Wilson y estoy sumamente preocupado por su reconocimiento como tal".[24]

Después de la Segunda Guerra Mundial, Hubble se dedicaría principalmente a asesorar y promover la construcción del telescopio de 5 metros en Monte Palomar.[25] Esta nueva herramienta astronómica costó seis millones de dólares de la época (1948), una cantidad inmensa en comparación con los gastos anteriores en astronomía, aunque irrisoria si se compara con los megaproyectos actuales.[26] Ante la pregunta de qué esperaban descubrir con el nuevo telescopio, tuvo el acierto de indicar que esperaban encontrar algo inesperado, y que si se supiera lo que se iba a descubrir no tendría utilidad construirlo;[27] algo diferente al discurso en boga actual, que requiere que se clarifique exactamente lo que se espera descubrir

---

[22] López Corredoira, 2008.
[23] López Corredoira, 2008.
[24] "I consider the velocity-distance relation, its formulation, testing and confirmation, as a Mount Willson contribution and I am deeply concerned in its recognition as such" (Block, 2012).
[25] Hubble, 1947; Mayall, 1970.
[26] Los seis millones de dólares norteamericanos en 1948 equivalen teniendo en cuenta la inflación económica a unos sesenta millones de dólares en 2019. Cantidades mucho mayores se dedican hoy en día a la construcción de grandes instrumentos de este tipo: el telescopio europeo E-ELT de 39 metros que está en construcción en Chile y se planea concluir en 2024 tiene un coste presupuestado de unos mil millones de euros, más el mantenimiento. Estimando el tiempo de vida del telescopio, supone un gasto de unos 100 millones de euros al año (cada segundo de operación con el telescopio cuesta unos diez euros); el telescopio espacial James Webb, que se prevé lanzar al espacio en 2021, tiene un coste presupuestado de unos diez mil millones de dólares norteamericanos, más costes de operación desde Tierra del orden de centenares de millones al año.
[27] Mayall, 1970; Narlikar, 2000.

cada vez que se presenta un nuevo proyecto multimillonario para nuevos colectores de fotones.

Si nos hemos de preguntar pues cuál ha sido la cosmovisión de este reputado científico del siglo XX, podemos concluir que, aparte de sus ideas científicas de alcance cosmológico, su visión ha sido la del típico pensador pragmático norteamericano con formación de abogado que convierte la ciencia en una gran empresa y la hipótesis científica en una verdad conveniente que merece defenderse y arrogarse, aunque en el fondo se sepa que no son exclusivas ni originales tales razones y que hay también razones para defender lo contrario.

Martín López Corredoira (1970) es Dr. en Cc. Físicas y Dr. en Filosofía, actualmente investigador titular en el Instituto de Astrofísica de Canarias, especializado en el área de galaxias y cosmología. Autor de alrededor de un centenar de artículos en revistas científicas internacionales con árbitro, la mitad de ellos como primer autor; unos 40 artículos de filosofía y diversos libros; los últimos: *The Twilight of the Scientific Age* (2013, en inglés), *Voluntad. La fuerza heroica que arrastra la vida* (2015).